\pdfoutput=1
\documentclass[fleqn,usenatbib]{mnras}

\volume{{\rm ArXiV post-print, 2019/05/23}}

\usepackage{newtxtext,newtxmath}

\usepackage[T1]{fontenc}
\usepackage{ae,aecompl}
\usepackage{subfig}
\usepackage{multirow}
\usepackage{siunitx}

\usepackage{graphicx}	
\usepackage{amsmath}	
\usepackage{amssymb}	
\usepackage[export]{adjustbox}

\title[PPPP with artificial Neural Networks]{Projected Pupil Plane Pattern (PPPP) with artificial Neural Networks}

\author[Huizhe Yang et al.]{
Huizhe Yang$^{1}$,
Carlos Gonzalez Gutierrez$^{2}$,
Nazim A. Bharmal$^{1}${\thanks{E-mail: n.a.bharmal@durham.ac.uk}}
\newauthor{
and F. J. de Cos Juez$^{2}$
}
\\
$^{1}$Centre for Advanced Instrumentation, Department of Physics, University of Durham, South Road, Durham DH1 3LE, UK\\
$^{2}$Mining Exploitation and Prospecting Department, C/Independencia n13, University of Oviedo, 33004 Oviedo, Spain}

\date{Accepted 2019 May 11, Published 2019 May 22, Post-print 2019 May 23}

\pubyear{2019}

\begin{document}

\hypersetup{pdfauthor={N.A.Bharmal on behalf of H.Y., C.G.G., and F.J.de C.J.},
            pdftitle={PPPP with artificial Neural Networks},
            pdfkeywords={atmosphere, turbulence, adaptive optics, ao, wavefront sensing, laser, lgs, ann, neural network, cnn},
            bookmarksnumbered=true}

\label{firstpage}
\pagerange{\pageref{firstpage}--\pageref{lastpage}}
\maketitle

\begin{abstract}
Focus anisoplanatism is a significant measurement error when using one single laser guide star (LGS) in an Adaptive Optics (AO) system, especially for the next generation of extremely large telescopes. An alternative LGS configuration, called Projected Pupil Plane Pattern (PPPP) solves this problem by launching a collimated laser beam across the full pupil of the telescope. If using a linear, modal reconstructor, the high laser power requirement ($\sim1000\,\mbox{W}$) renders PPPP uncompetitive with Laser Tomography AO. This work discusses easing the laser power requirements by using an artificial Neural Network (NN) as a non-linear reconstructor. We find that the non-linear NN reduces the required measurement signal-to-noise ratio (SNR) significantly to reduce PPPP laser power requirements to $\sim200\,\mbox{W}$ for useful residual wavefront error (WFE). At this power level, the WFE becomes 160\,nm root mean square (RMS) and 125\,nm RMS when $r_0=0.098$\,m and $0.171$\,m respectively for turbulence profiles which are representative of conditions at the ESO Paranal observatory. In addition, it is shown that as a non-linear reconstructor, a NN can perform useful wavefront sensing using a beam-profile from one height as the input instead of the two profiles required as a minimum by the linear reconstructor.

\end{abstract}




\section{Introduction}
Adaptive Optics is a technology which corrects for the aberrations introduced by turbulence in the atmosphere and so improves the quality of the point spread function (PSF) for ground-based astronomical observations \citep{Hardy}. It becomes more important as the telescope diameter increases but the availability of stars with minimum brightness requirements to act as guide stars for wavefront sensor (WFS) measurements is independent of diameter. To overcome the lack of natural guide stars (NGS), laser guide stars (LGS) are deployed \citep{Fugate} as a partial substitute. A LGS is created by a laser projected to form a compact beacon in the atmosphere and the light scattered back to the telescope is analysed with a dedicated WFS. For a LGS AO system, the key disadvantage is that high altitude turbulence is illuminated by the LGS over a smaller region than the illumination from the scientific target because of the finite LGS altitude: the focus anisoplanatism problem \citep{Hardy}. The wavefront error (WFE) from focus anisoplanatism becomes more pronounced for larger telescopes \citep{Fried}: $\sim$155\,nm RMS if only one sodium LGS ($\sim90$\,km altitude) is used on a 10\,m telescope as described by \cite{Keck}, and $\gtrsim300\,\mbox{nm}$ for 30\,m-scale ELTs. \\
Laser Tomography AO (LTAO) is the conventional solution to focus anisoplanatism by using several LGSs generated at different positions in the sky and then estimating the 3D turbulence \citep{TF90}. An alternative to LTAO is Projected Pupil Plane Pattern (PPPP) which avoids the multiple-LGS-and-tomography solution, and so has several unique features compared to LTAO. First, turbulence is sensed during the projection of the laser beam, which is not focused but instead collimated and from the telescope primary mirror. This sampling of the atmospheric volume, equivalent to that illuminated by the target, by the laser beam is how the focus anisoplanatism is eliminated. Second, no WFS is required but instead a camera is used to image the back-scattered light as a beam-profile, implying an estimate of the beam-profile after propagation for a certain distance is the measurement. Third, the wavefront reconstruction does not employ tomography nor require \textit{a priori} knowledge of the turbulence profile. PPPP has been demonstrated as an effective solution to focus anisoplanatism using simulation and laboratory experiments, \cite{Yang} and \cite{Yang2} respectively, and can achieve equivalent performance to a Shack-Hartmann (SH) WFS using a natural guide star. However, the linear reconstruction method used so far performs poorly when including photon noise (dependent on laser power) and detector read noise.  According to \cite{Yang}, a $\sim$1000\,W laser is required for PPPP to reach similar performance to the use of a SH WFS with a sodium LGS. To advance PPPP to the level of a practical alternative to LTAO, this work concentrates on using a nonlinear reconstruction method, an artificial Neural Network (NN), to significantly reduce laser power requirements while retaining a useful WFE.\\
A NN is a machine learning-based algorithm which has the ability to learn from different examples and extrapolate that knowledge to unseen data. They were traditionally inspired by human neurons \citep{Rosenblatt}, but have been developed to form the Deep Learning models widely used today \citep{LeCun}. Neural Networks have been used with AO successfully on-sky, including recently to produce a tomographic reconstructor operating with multiple WFSs using an asterism of guide stars as described by \cite{James}. However, each potential asterism demands a different NN algorithm which in turns leads to a set of time-consuming training processes. In contrast, applying the NN methodology for PPPP has the advantage that the laser beam is under control and so can be fixed: once trained a NN-based reconstructor needs not necessarily be retrained when changing the telescope pointing direction.\\
The layout of this paper is as follow. In section \ref{sec:PPPP theory and linear reconstruction} we review the PPPP theory and its conventional, linear reconstruction method. In section \ref{sec:Neural Network reconstruction} the neural network reconstruction methodology is described by using convolutional neural networks. In section \ref{sec:Results} simulation results are presented, comparing the performance of a NN-based reconstructor with that from using the linear reconstruction, and both against a NGS SH WFS which is the baseline we aim for. In section \ref{sec:Conclusion} we draw our conclusions.

\section{PPPP theory and linear reconstruction}
\label{sec:PPPP theory and linear reconstruction}
\subsection{PPPP theory}
\label{sec:PPPP theory}

\begin{figure}
   \begin{center}
   \includegraphics[scale=0.25]{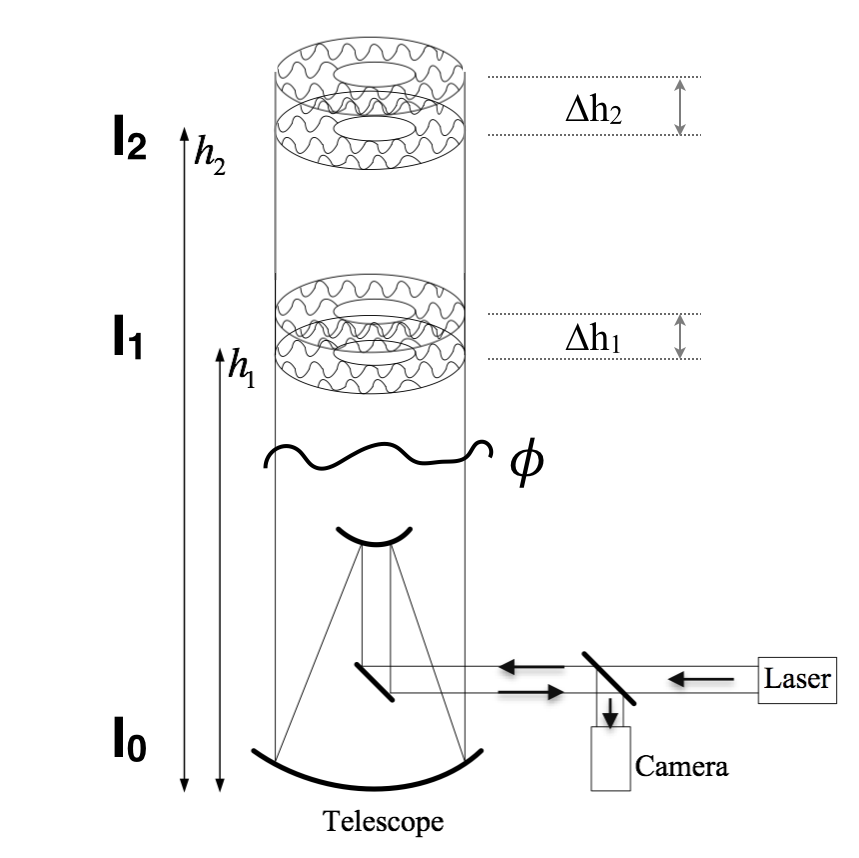}
    \caption{A schematic of how PPPP could be implemented. A collimated laser beam is propagated upward into the atmosphere from the whole telescope primary mirror, and encounters aberrations, $\phi$. Light is back-scattered from an altitude, $h$, is recorded to form $I_1$ when $h$ is in the range $h_1\pm\Delta{h_1}/2$ and similarly $I_2$ when $h$ is in the range $h_2\pm\Delta{h_2}/2$. 
    }
    \label{fig:Schematic diagram of PPPP}
   \end{center}
\end{figure}

PPPP is reminiscent of a Curvature WFS \citep{Roddier}, since both are based on the relationship between phase and intensity by the transport-of-intensity equation (TIE),

\begin{equation}
      \label{eq:TIE}
      k\partial_z I=-\nabla\cdot \big (I\nabla\phi\big),\\
   \end{equation}
   which can be approximated as,
   \begin{equation}
       k\frac{I_2-I_1}{h_2-h_1}=-\nabla\cdot \big (I_0\nabla\phi\big)=-\nabla I_0\cdot\nabla\phi-I_0\nabla^2\phi,
      \label{eq:approx TIE}
\end{equation}
where $I_0$, $I_1$ and $I_2$ are the intensity patterns at the propagation distances of zero, $h_1>0$ and $h_2>h_1$ respectively. The aberration $\phi$ can be located at $z\leq{}h_1$. From $I_0$, $I_1$ and $I_2$, we can retrieve the phase $\phi$, except its mean (piston), according to equation (\ref{eq:approx TIE}). A PPPP schematic diagram is shown in Fig. \ref{fig:Schematic diagram of PPPP}. A laser beam is expanded to fill the pupil of the telescope and propagates as a collimated beam upward through the atmosphere. When the laser pulse reaches an altitude of $h_1$, the light that then back-scatters to the surface is used to form an image of the beam-profile within the range gate $\Delta h_1$ using a camera conjugate to $h_1$. This beam-profile, after calibration, is referred to in this work as $I_1$. When some time later the laser pulse reaches an altitude of $h_2$ and is similarly scattered back, a second image is taken with a camera conjugate at $h_2$ to obtain $I_2$ . With the obtained, calibrated beam-profiles $I_1$ and $I_2$, we can retrieve the turbulence wavefront $\phi/k$ via equation (\ref{eq:approx TIE}). To have a finite range gate depth, a pulsed laser is required.\\

The description of how PPPP operates can be divided into three processes: first, a collimated beam is propagated upwards through the atmospheric turbulence from the telescope pupil plane to two altitude bins--the upwards propagation; then the back-scattered light from those altitude ranges is recorded as a beam-profile using either the same telescope or one nearby--the return path; and finally, the estimation of the non-constant component of the aberrations encountered by using the subtraction of the beam-profiles---the reconstruction. A detailed description of how these processes were simulated together with results was presented in \cite{Yang}. It was demonstrated that the signal obtained using the PPPP method is generated during the upward propagation of the laser and that aberrations encountered in the return path can be neglected if the beam-profiles are measured at a sufficiently low angular sampling to allow seeing to be neglected.

\subsection{Linear reconstruction}
\label{sec:Linear reconstruction}

The reconstruction of phase, and therefore the wavefront, is by decomposition into Zernike polynomials, where $Z_j$ is the $j$-th Zernike polynomial \citep{Noll}. The reconstructed phase $\hat{\phi}$ is then decomposed as
   \[
      \hat{\phi}=\sum_{i=2}^{N_z+1}a_iZ_i(r,\theta),
   \]
if $N_z$ Zernike polynomials are used for reconstruction (excluding piston).\\
The linear reconstruction algorithm was proposed by \cite{Gureyev}. When the intensity distribution $I_0$ in equation (\ref{eq:TIE}) is smoothly approaching zero on the boundary of the pupil, then it is possible to introduce the matrix $\mathbf{M}$ with its elements defined as

   \begin{equation}
       \label{eq:M}
       M_{ij}=\int_{0}^{2\pi}\int_{0}^{R}I_0\nabla Z_i \cdot \nabla Z_jrdrd\theta,
   \end{equation}
where $R$ is the radius of the pupil. Given $\mathbf{M}$, the reconstructed Zernike coefficient vector $\vec{a}$ equals

   \begin{equation}
       \label{eq:recon}
       a_j=R^2\mathbf{M}^{-1}F_j\quad \text{or} \quad \vec{a}=R^2\mathbf{M}^{-1}\vec{F},
   \end{equation}
where
   \begin{equation}
       F_j = kR^{-2}
           \int_{0}^{2\pi}\int_{0}^{R}
             \frac{I_2-I_1}{h_2-h_1}Z_{j}\;r\,dr\,d\theta.
   \label{eq:Fj}
   \end{equation}

Given $I_0$, $\mathbf{M}^{-1}$ can be formed in advance and then calculating $F_j$ requires only a difference of the measured intensities, $I_1$ and $I_2$, and an integration weighted by $Z_j$. To be consistent with \cite{Yang},

   \begin{equation}
      \label{eq:I0}
      I_0=-0.1297+\exp{[-r^2/(2\times 1.05^2)]},
   \end{equation}
which satisfies the requirements \citep{Gureyev} for using equation (\ref{eq:M}). This Gaussian-like beam-profile is shown in Fig. \ref{fig:black_box}.

\section{Neural Network reconstruction}
\label{sec:Neural Network reconstruction}
Having discussed the methodology of the linear reconstructor, we now describe how a NN is developed to perform the equivalent task.  A characteristic of a NN is its inherent ability to generalise from {\em a priori} known set of inputs. By exposing a NN to these inputs, together with their desired outputs, the NN can predict an output when confronted with a superposition of a number of the independent training sets from combining a number of the synaptic pathways. Recently, the Convolutional Neural Network (CNN) and increased computation power have together been shown significant performance in different fields such as image classification \citep{krizhevsky}, object detection \citep{sermanet}, or speech recognition \citep{graves}.\\

\subsection{Convolutional Neural Networks}
\label{sec:Convolutional Neural Networks}
\begin{figure*}
      \begin{center}
      \includegraphics[scale=0.6]{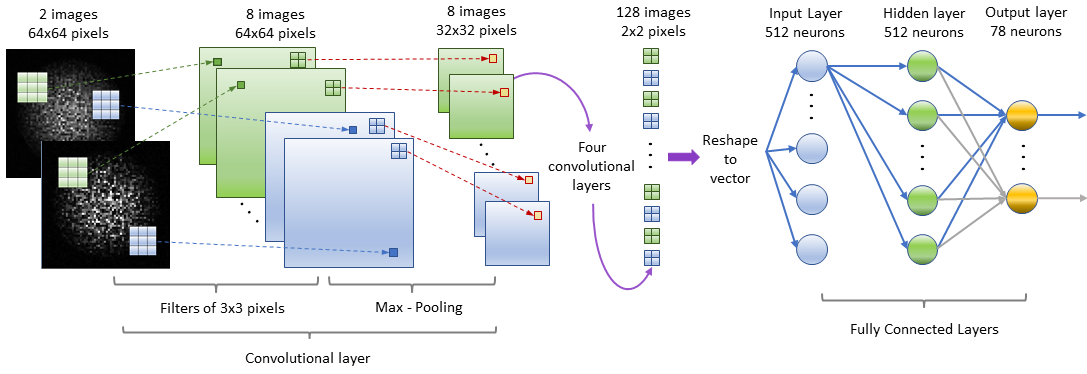}
       \caption{Convolutional Neural Network Diagram. The two input images are convolved by 4 filters of 3x3 pixels, creating 64 images. After this pooling is applied to each image, the image size is reduced by half. This process is repeated four times but using only 2 filters per layer, creating 128 images of 2x2 pixels which are flattened and passed as inputs to the fully connected stage. At the end the 78 Zernike coefficients are returned as the output.
       }
       \label{fig:Convolutional Neural Network}
      \end{center}
   \end{figure*}

A Neural Network is composed by several layers of neurons, connected to each other in a feed-forward fashion. All the connections between neurons are called weights. As a sub-type of NNs, CNNs are characterized by the appearance of convolutional layers, which help in the extraction of features from an image. These layers are composed of several filters that are convolved with the input image, therefore generating a new set of processed images as outlined in Fig. \ref{fig:Convolutional Neural Network}. After the convolution, an activation function is applied. There are several types of functions, such as sigmoid or hyperbolic tangent, although the most common type is the Rectified Linear Unit (ReLU) \citep{LeCun}. It is common to use a pooling layer after the activation function, which reduces the size of the produced images, by extracting the maximum or median value from a certain region of pixels. This set of layers could be nested several times and will reduce the size of the processed input image, while increasing the number of processed images. The last stages of a CNN are fully connected layers that connect all the neurons from one layer to those in the next layer. These fully connected layers can be repeated as required, and the CNN ends with an output layer. To connect the convolutional stage with the fully connected layers, it is necessary to reduce the dimensionality of the images. This reduction is achieved by flattening the outputs of the convolution into a 1D vector. A summary of a complete CNN used for PPPP wavefront reconstruction is shown in Figure \ref{fig:Convolutional Neural Network}.\\

A key stage in obtaining a usable NN is the learning, or training, process. By using a dataset of known inputs and associated outputs, it is possible to calculate optimal values for the weights. Initially, the weights are random and an input is propagated through the network. The output of the CNN is computed and compared with the expected output, which results in a residual error. This error is back-propagated \citep{rumelhart} through the network and the weights are updated accordingly. By iterating through the dataset, this process is repeated and the weights are updated until the final input data has been seen: this process is called an epoch. Training is ended after a certain number of epochs when some suitable criterion to evaluate the network has been met, we discuss this further below. The CNN architecture has been demonstrated as particularly advantageous for image processing, and since the input are two beam-profiles, $I_1$ and $I_2$, it is appropriate to be used in this work and referred as NN in the following.

\subsection{NN implementation for PPPP}
\label{sec:CNN Architecture for PPPP}

\subsubsection{NN Parameters}
\label{sec:NN Parameters}
\begin{figure}
     \begin{center}
      \includegraphics[trim={12cm 2cm 10cm 2cm}, clip,scale=0.45]{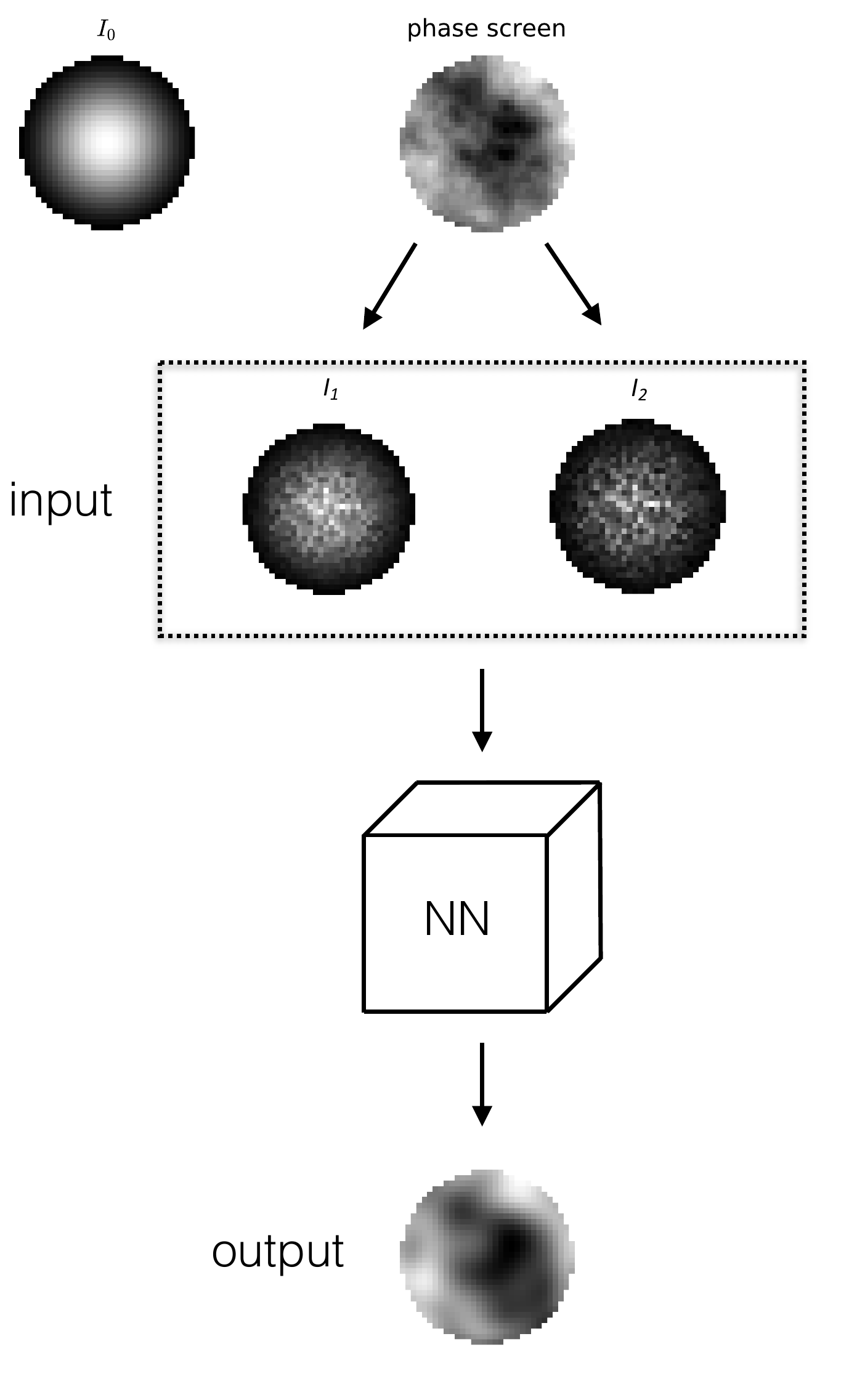}
       \caption{Schematic diagram of the generation of PPPP signal and NN reconstructor as a black box. A Gaussian-like beam at the pupil $I_0$ propagates through a random phase screen to $h_1$ and $h_2$, forming images $I_1$ and $I_2$ respectively. The input for NN reconstructor then is the two images $I_1$ and $I_2$ and the output is the reconstructed 78 Zernike coefficients (here shown as the reconstructed phase for convenience). }
       \label{fig:black_box}
      \end{center}
   \end{figure}
For PPPP, if we describe a NN as a ``black box" non-linear reconstructor, as shown in Fig.~\ref{fig:black_box}, then its inputs are two images of the scattered intensity patterns--beam-profiles--from two different altitudes, i.e.~$I_1$ and $I_2$. The expected output is a vector of 78 Zernike coefficients representing the reconstructed wavefront.\\
The NN reconstructor is composed of five convolutional layers and two fully connected layers at the end. The inputs, $I_1$ and $I_2$, have dimensions of $64\times64$ pixels. The outputs of the reconstructor is $\vec{a'}$, the first 78 Zernike coefficients but scaled so their expected values, with respect to the training dataset, are individually in the range $\pm1$. This normalization improves performance by {\em a priori} accounting for the large difference in expected range between high and low order polynomials.\\
The NN is created using the {\em TensorFlow} \citep{TensorFlow} software. Each convolutional layer applies 2 filters of $3\times3$ pixels to all images (except for the first layer, where 4 filters are applied), followed by a max pooling function which reduces each image size by half. At the end of the convolutional stage, there are 128 images of $2\times2$ pixels, which are flattened into a 1D vector of size 512. This vector is the input to the first fully connected layer, composed of 512 neurons, followed by the final output layer which has 78 neurons. The activation function type is the Leaky ReLU \citep{maas}, which improves the phase reconstruction by allowing non-zero values when the input is negative \citep{xu}.\\
For the training process, the Adagrad backpropagation algorithm \citep{duchi} has been used, along with a batch size of 128 samples. The error of the backpropagation is calculated using the Root Mean Squared Error (RMSE). Initial experimentation suggests a learning rate of 0.01 as a good compromise between the speed and correct outputs. Initial values of the weights are set by Xavier initialization \citep{glorot} which has been demonstrated an improvement over the Gaussian initialization.

\subsubsection{Training dataset}
\label{sec:Dataset}
During the training process it is necessary to expose the NN to a large number of pairs of inputs and desired outputs. This training dataset should cover the full range of possible scenarios, and previous experiments in atmospheric wavefront reconstruction show that a NN can accurately predict an output when trained with a superposition of independent training sets \citep{James}. The conclusion is that not every possible turbulent profile is required but instead a basis set is sufficient for training. Such as basis set for PPPP is now described.\\
Table~\ref{tab:PPPP parameters} shows the parameters used to generate the training dataset from the PPPP model in an AO simulation platform $Soapy$ \citep{Andrew} including the upward propagation, return path and reconstruction according to \cite{Yang}. The tip/tilt modes are excluded from both input phase screens and reconstructed Zernike coefficients because the tip-tilt signal, a global movement of the beam-profile, is also affected by the return path. As with the use of a LGS, it is necessary to use a NGS to provide the tip-tilt information. The parameters are chosen to balance the PPPP performance and complexity. Four sets of training data were created, each with a constant laser power: 10, 20, 200\,W and infinite power (photon noise-free). For each power simulated, 100 altitudes for one turbulence layer, $h$,  distributed between 0 and 10\,km are defined, with 10 values of $r_0$ between 0.08\,m and 0.28\,m per turbulence layer altitude and 300 realizations of a random phase-screen for each $r_0$ value. Thus for each turbulence altitude, there would be 3000 pairs of beam-profiles for training. This leads to 300,000 pairs of beam-profiles for each laser power including 100 turbulence layer altitudes, with each pair created from a well-defined $r_0$ and $h$ value: this is the basis set. These data can be used to train four different Neural Networks, each for a specific laser power, or used together to train one combined Neural Network which is laser power agnostic.

\begin{table}
	\centering
	\caption{PPPP parameters for training dataset. $D$ is the telescope diameter. The number of pixels across the selected square to pad the pupil is $N_{total}$, to reduce edge effects during propagation, and $N_{pupil}$ is the number of illuminated pixels across the pupil}. The transmission of the optical components is $T_0$, and $T_A$ is the one-way transmission of the atmosphere. The outer and inner scale are $L_0$ and $l_0$, respectively. The laser pulse length is $\Delta h_1$ and $\Delta h_2$ for $h_1$ and $h_2$ respectively. $I$ is average laser power and $\eta$ is the quantum efficiency of photon detector.
	\label{tab:PPPP parameters}
    \begin{tabular}{ll}
    \hline
  \multicolumn{1}{l}{simulation} & \multicolumn{1}{l}{turbulence} \\
  \hline
  $D$=4\,m & one turbulence layer \\
  $h_1$=10\,km & altitude: 0 to 10\,km  \\
  $h_2$=20\,km & $r_0$: 0.08 to 0.28\,m (at 500nm) \\
   $N_{total}$=64 & $T_0$=0.5; $T_A$=1 \\
   78 Zernike modes& $L_0$=100\,m; $l_0$=0.01\,m \\
  Gaussian-like beam $I_0$ &  \\
  \hline
  \multicolumn{1}{l}{laser} & \multicolumn{1}{l}{camera} \\
  \hline
  $\lambda$=1064\,nm & $N_{pupil}$=54 \\
  $\Delta h_1$=1\,km; $\Delta h_2$=5\,km & $\eta$=0.8 \\
  $I$ (W): 10, 20, 200 \& infinite & read noise: 3$e^-$ \\
  laser frequency: 5\,KHz & exposure time: 2.5\,ms \\
  \hline
\end{tabular}
\end{table}

\subsubsection{Discussion}
\label{sec:cnn:pppp:discussion}
During training, it is necessary to use early-stopping techniques to avoid overfitting the Neural Network to the training data. Within several thousands of epochs, the computed RMSE used to backpropagate through the NN kept decreasing. However the residual wavefront error (WFE) from an AO simulation (see the next section for details) using the NN models of different numbers of epochs stops decreasing. This occurs when the NN is overfitted to a specific laser power, and the simulation reveals this by a corresponding increase in WFE for other laser powers. It is evident that the reconstruction quality deteriorates when overfitting. To avoid overfitting the training process is therefore ended after 1000 epochs.\\
One of the advantages of NNs, is the flexibility of their topology (internal configuration of layers) when altering the size of the input or output. If such an alteration is performed then usually it is only necessary to re-train the network rather than having to redesign its topology. Although the linear reconstructor requires at least two beam-profiles as inputs, the NN can be adapted and then trained with only one beam-profile from one altitude (see section \ref{sec:res:1bp} for details). For this specific case, the modifications are only to the first convolutional layer which then has 8 filters instead of 4. The other parameters and architecture are unchanged.\\ 

\section{Results}
\label{sec:Results}

\subsection{Suitability for real-time operation}

The number of operations for each reconstruction method is now discussed to highlight suitability for real-time use. For the NN, reconstruction is calculated network-layer by layer. In the convolutional stage, each image is multiplied with all the filters. The amount of calculations required for each subsequent convolutional layer is reduced substantially when propagating through the NN hence the convolution operations dominate. In the fully connected layers the number of operations is equal to the product of the number of input neurons by the number of output neurons. The total number of arithmetic operations for the NN reconstruction is therefore estimated as $\sim875,000$. In comparison, the linear reconstructor uses a matrix vector multiply operation (the reconstruction matrix of size $N_Z^2$ is multiplied with measurement related vector $F$, which is a length $N_Z$ vector). This makes the matrix-vector-multiplications require a $O(N_Z^2)$ number of calculations. However, the formation of $F_j$ requires pixel-by-pixel processing for $N_p=\pi (N_{pupil}/2)^2\approx2300$ per beam-profile. This is $O(2N_Z{}N_p)$ for two beam-profiles and so dominates the number operations. It is estimated that $\sim365,000$ operations are required for the linear reconstruction. Therefore the NN is only $\sim$2 times more computationally complex than the linear method and the processing of the input data, $I_1$ and $I_2$, dominates in both methods.

\subsection{Validation of the NN reconstructor}
\label{sec:Validation of the NN reconstructor}
\begin{figure}
	\begin{centering}
	\includegraphics[scale=0.5]{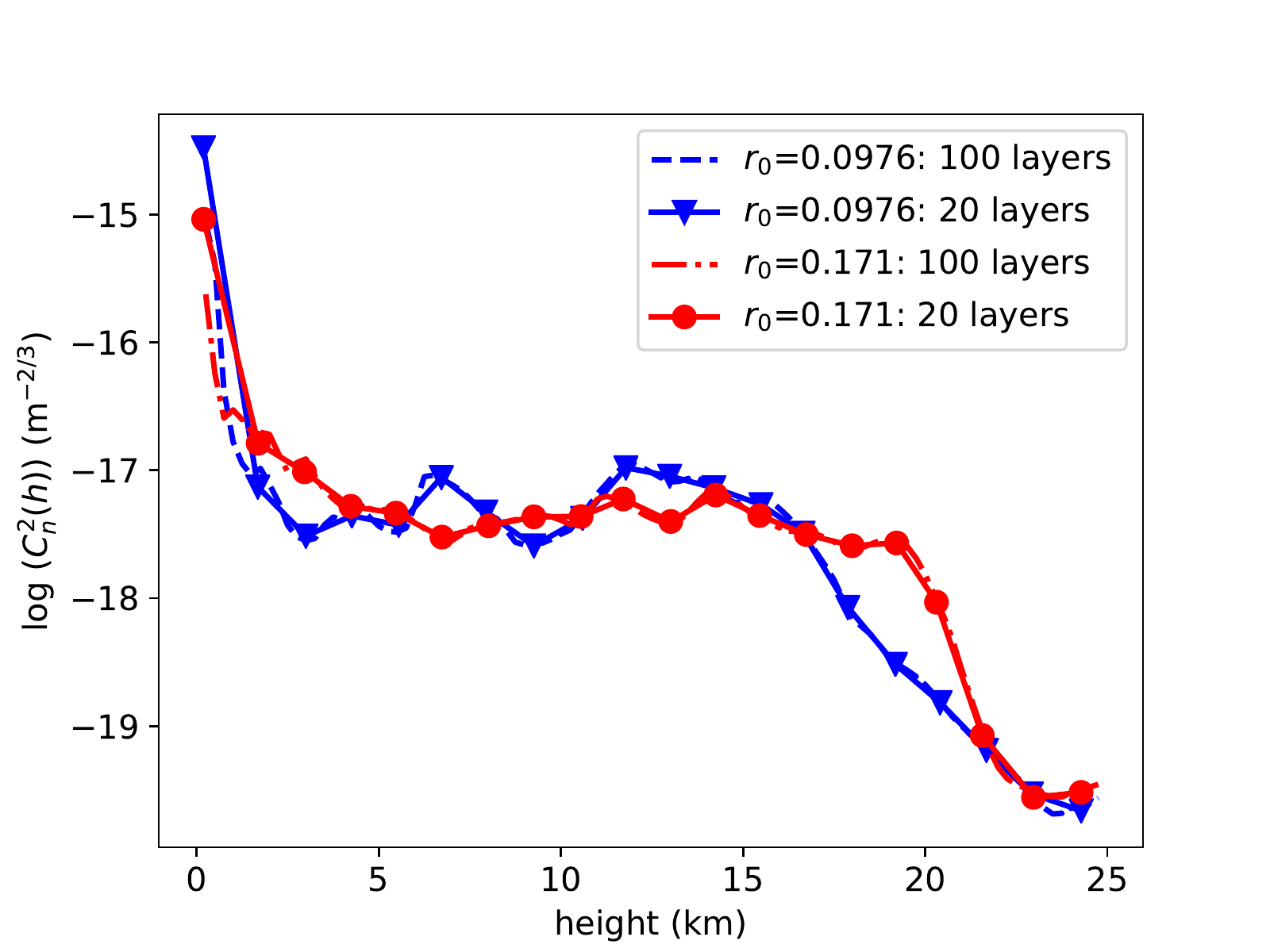}
 	\caption[Two representative optical turbulence profiles measured at ESO Paranal]{Two representative optical turbulence profiles measured at ESO Paranal from \cite{Ollie} with $r_0$'s equalling 0.0976 and 0.171\,m. They have both 100- and 20- turbulence layer representations; the 20 layer representation in used in this work.}
    \label{fig:cn2}
	\end{centering}
\end{figure}

\begin{figure}
    \begin{centering}
    \includegraphics[trim={2cm 14cm 2cm 3.7cm},clip,scale=0.5]{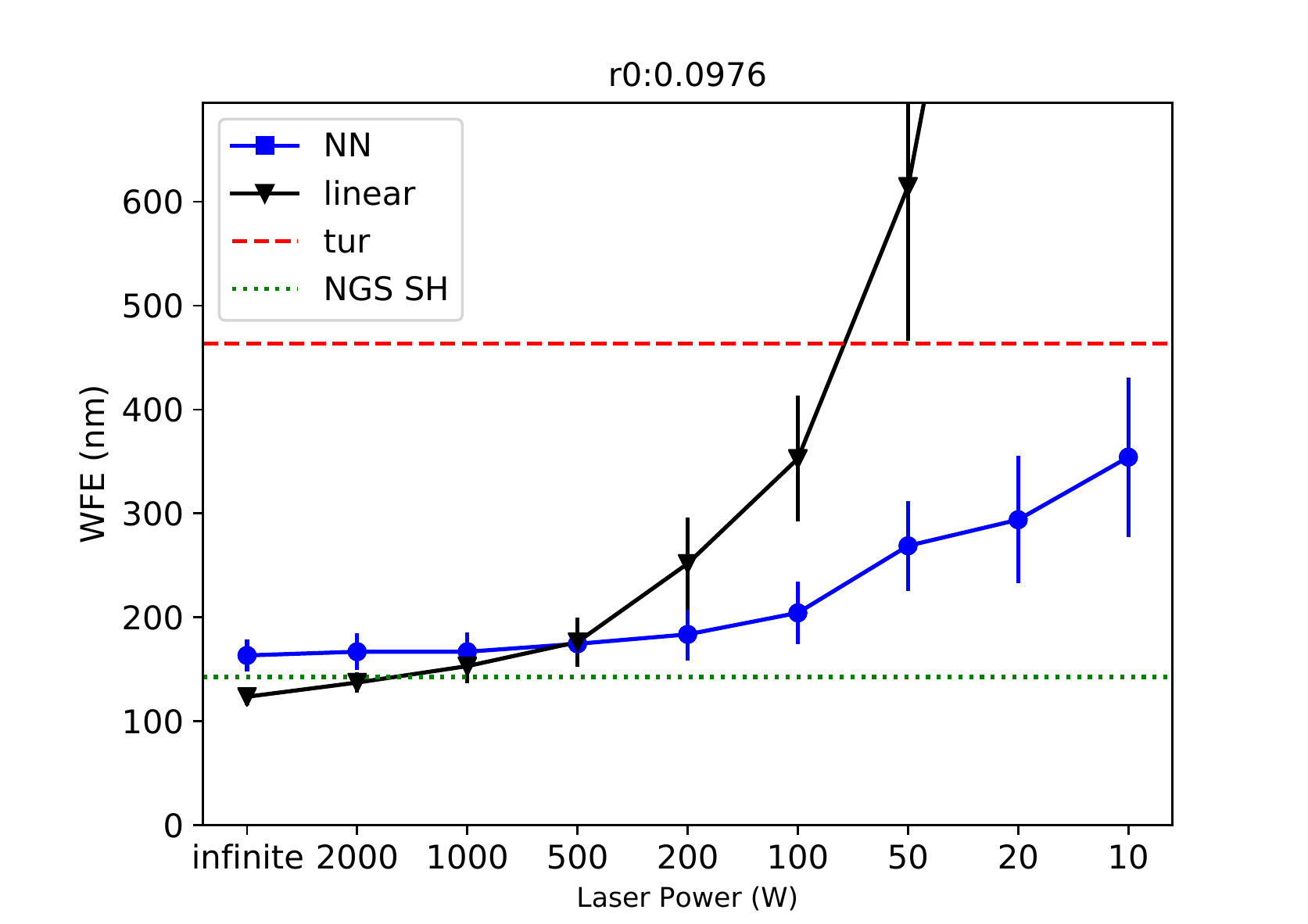}\\
    \includegraphics[trim={2cm 14cm 2cm 3.7cm},clip,scale=0.5]{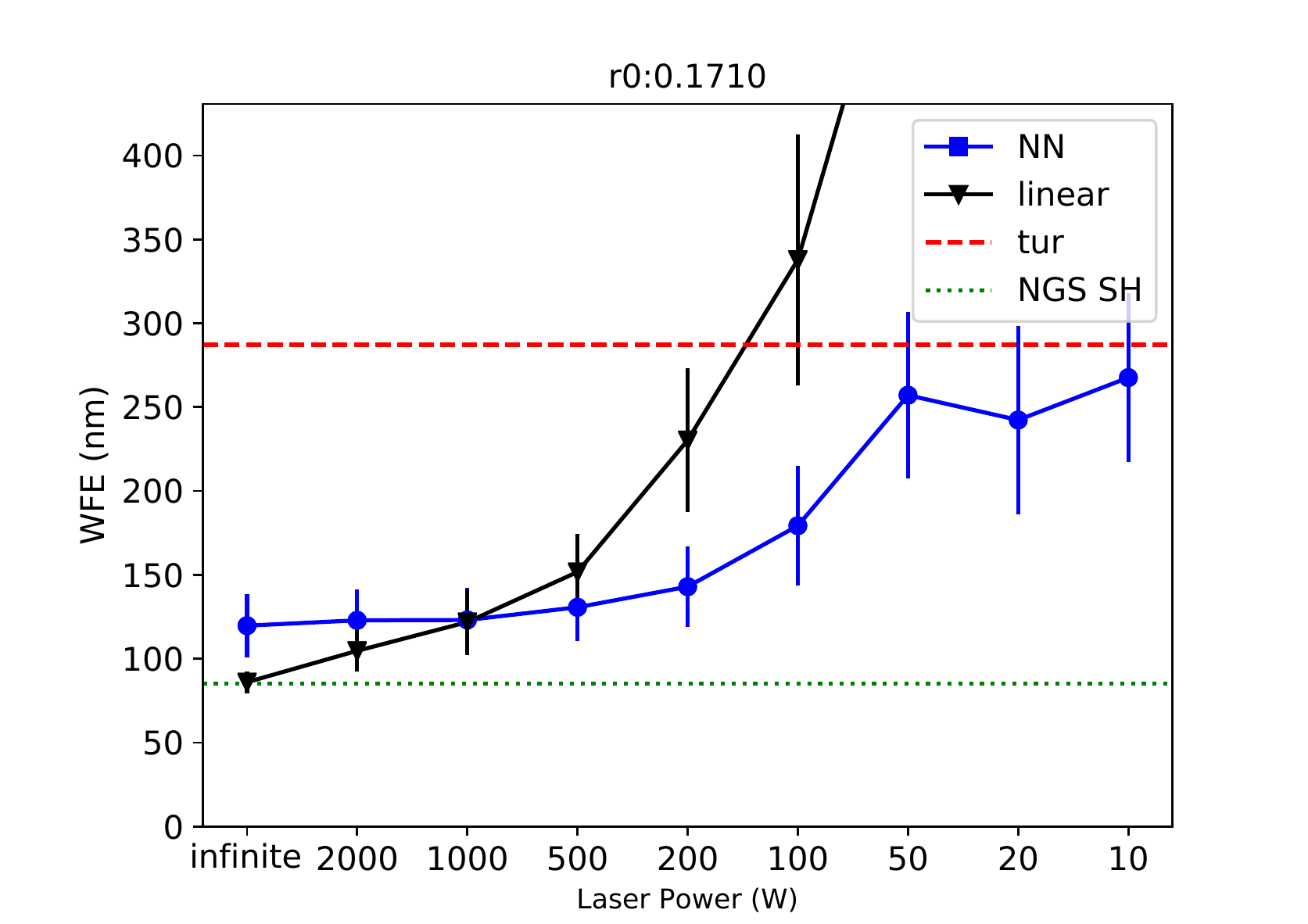}
   \caption{The WFE (nm) of linear and NN reconstructor with different laser powers. The ``NGS SH'' shows the ideal performance and the ``tur'' represents the RMS of the uncorrected wavefront. The result is an average of 50 random turbulence realizations from the $Soapy$ simulations.}
    \label{fig:WFE results}
    \end{centering}
\end{figure}

For the validation two representative optical turbulence profiles measured at Cerro Paranal \citep{Ollie} are used, with $r_0$ equalling 0.0976 and 0.171\,m respectively. The turbulence profiles are shown in Fig.~\ref{fig:cn2}. Both the profiles are consistent with long-term statistical analysis \citep{James2018}: turbulence at the ground is dominant and there are several peaks between 5 and 20\,km. Comparing these two profiles, we find that there is a stronger ground layer for the $r_0=0.0976$ profile, while the turbulence layer around 20\,km is stronger for the $r_0=0.171$\,m profile. These two profiles were chosen since their $r_0$ values cover the worst and best seeing from 83 nights in ESO Paranal, with the probability percentage of their occurrence equalling 9.7\% and 1.4\%.\\
To validate the NN reconstructor, it was included into $Soapy$, which is a
Monte-Carlo Adaptive Optics simulation written in the Python programming
language \citep{Andrew}, with an integrated PPPP simulation model \citep{Yang}.
For a wavefront sensing comparison with PPPP, a zero-noise SH WFS associated
with an infinitely bright NGS is implemented with $26 \times 26$ sub-apertures.
The synthetic DM in the simulation can reproduce, exactly, the first 78 Zernike
polynomials. The simulation is configured to run in open-loop.\\
Initially, we discuss the NN after it is trained with all laser powers. From the two turbulence profiles shown in Fig. \ref{fig:cn2} and the PPPP parameters listed in Table \ref{tab:PPPP parameters}, the average wavefront error is obtained from 50 random turbulence realization per profile. The results for different laser powers (varying photon noise in the measured beam-profiles) are shown in Fig.~\ref{fig:WFE results}. It is found that the NN reconstructor can significantly reduce the residual wavefront error when the laser power is less than 1000\,W, which in turn reduces the laser power requirements for implementation. As expected, for both the linear and NN reconstructors, with larger WFE the corresponding standard deviation increases. Comparing $r_0=0.0976$\,m and $r_0=0.171$\,m, we find that the intersection of the two reconstructors is around 500\,W for $r_0=0.0976$\,m, and 1000\,W for $r_0=0.171$\,m. It means that the linear reconstructor performs better, relatively, for poorer seeing (smaller $r_0$). For the linear reconstructor, equation (\ref{eq:approx TIE}) implies the signal, proportional to $I_2-I_1$, is a linear function of the phase and therefore larger for poorer seeing. However the NN reconstructor is not as sensitive to the seeing which suggests that the NN is using $I_1$ and $I_2$ independently and not their difference directly. The intriguing suggestion is that sufficient information for reconstruction is contained within each beam-profile, and this is discussed further below.

\begin{figure*}
    \begin{centering}
    \subfloat[][\centering $r_0$=0.0976\,m; infinite power]
{	
   \includegraphics[trim={2cm 17cm 2cm 0cm},clip,scale=0.5]{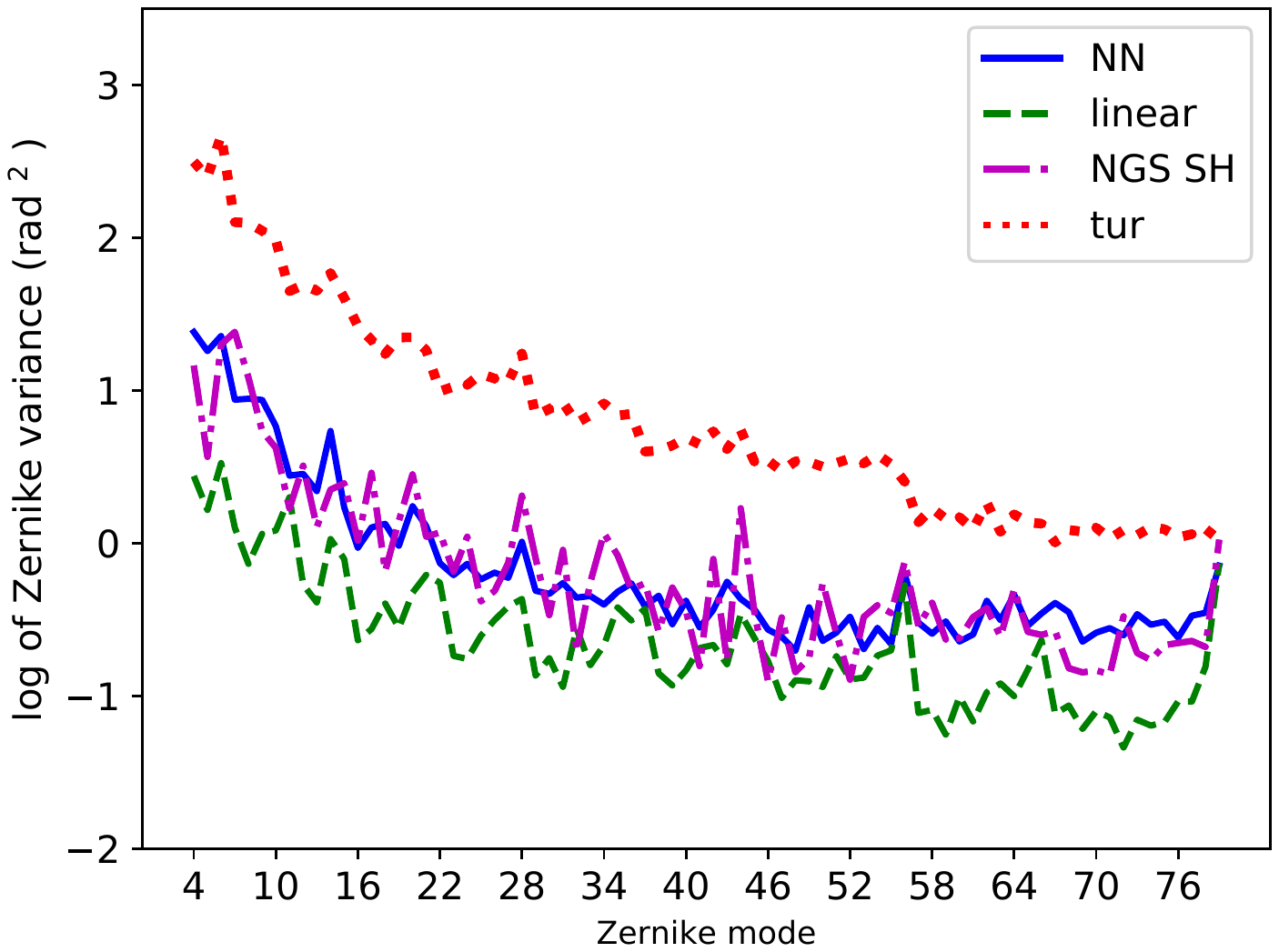}
    \label{fig:Zer_r0_006_inf}
}
\subfloat[][\centering $r_0$=0.171\,m; infinite power]
{	
   \includegraphics[trim={2cm 17cm 2cm 0cm},clip,scale=0.5]{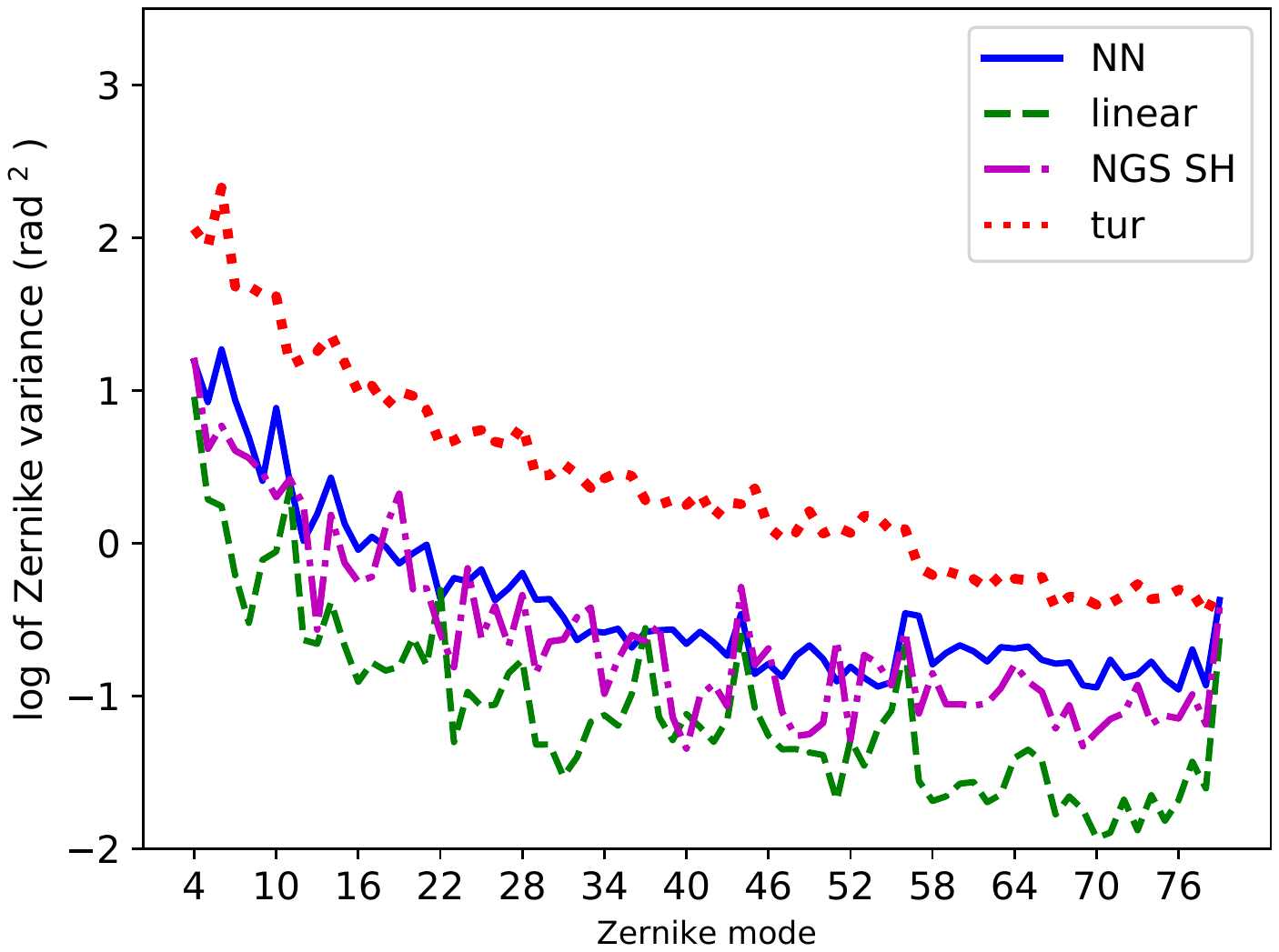}
    \label{fig:Zer_r0_025_inf}
}\\

\subfloat[][\centering $r_0$=0.0976\,m; 200\,W]
{	
   \includegraphics[trim={2cm 17cm 2cm 0cm},clip,scale=0.5]{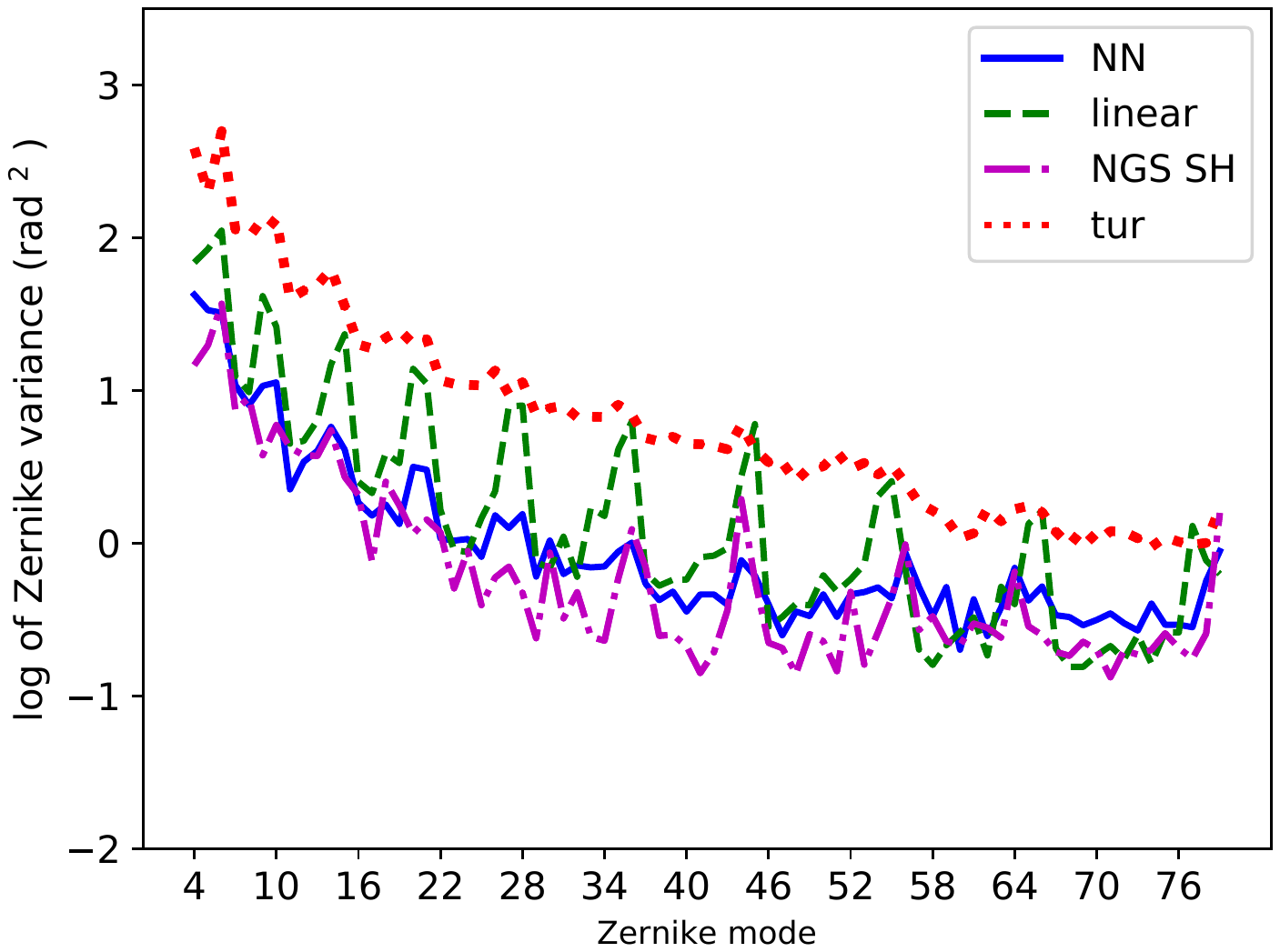}
    \label{fig:Zer_r0_006_200W}
}
\subfloat[][\centering $r_0$=0.171\,m; 200\,W]
{	
   \includegraphics[trim={2cm 17cm 2cm 0cm},clip,scale=0.5]{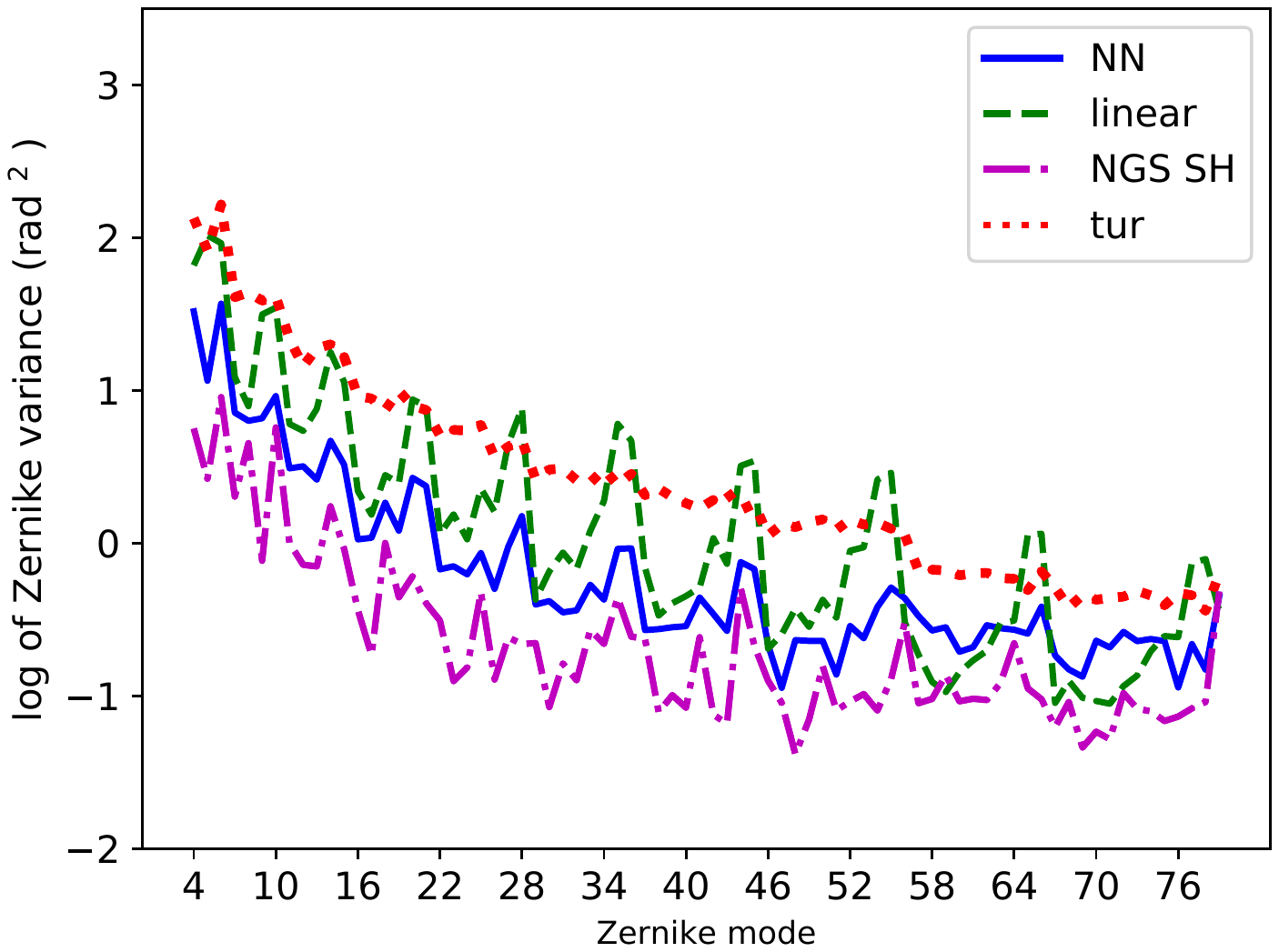}
    \label{fig:Zer_r0_025_200W}
}\\
\subfloat[][\centering $r_0$=0.0976\,m; 20\,W]
{	
   \includegraphics[trim={2cm 17cm 2cm 0cm},clip,scale=0.5]{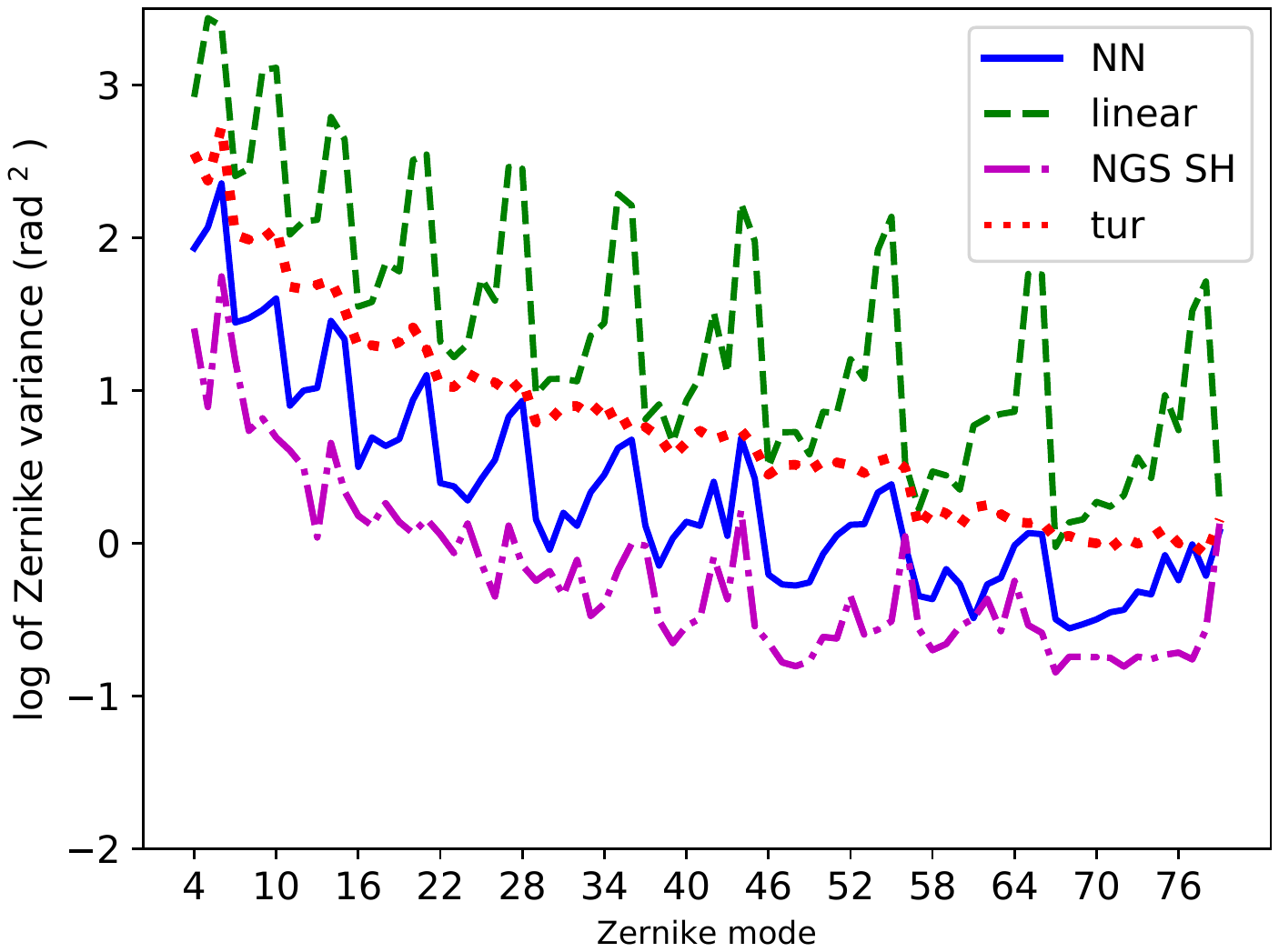}
    \label{fig:Zer_r0_006_20W}
}
\subfloat[][\centering $r_0$=0.171\,m; 20\,W]
{	
   \includegraphics[trim={2cm 17cm 2cm 0cm},clip,scale=0.5]{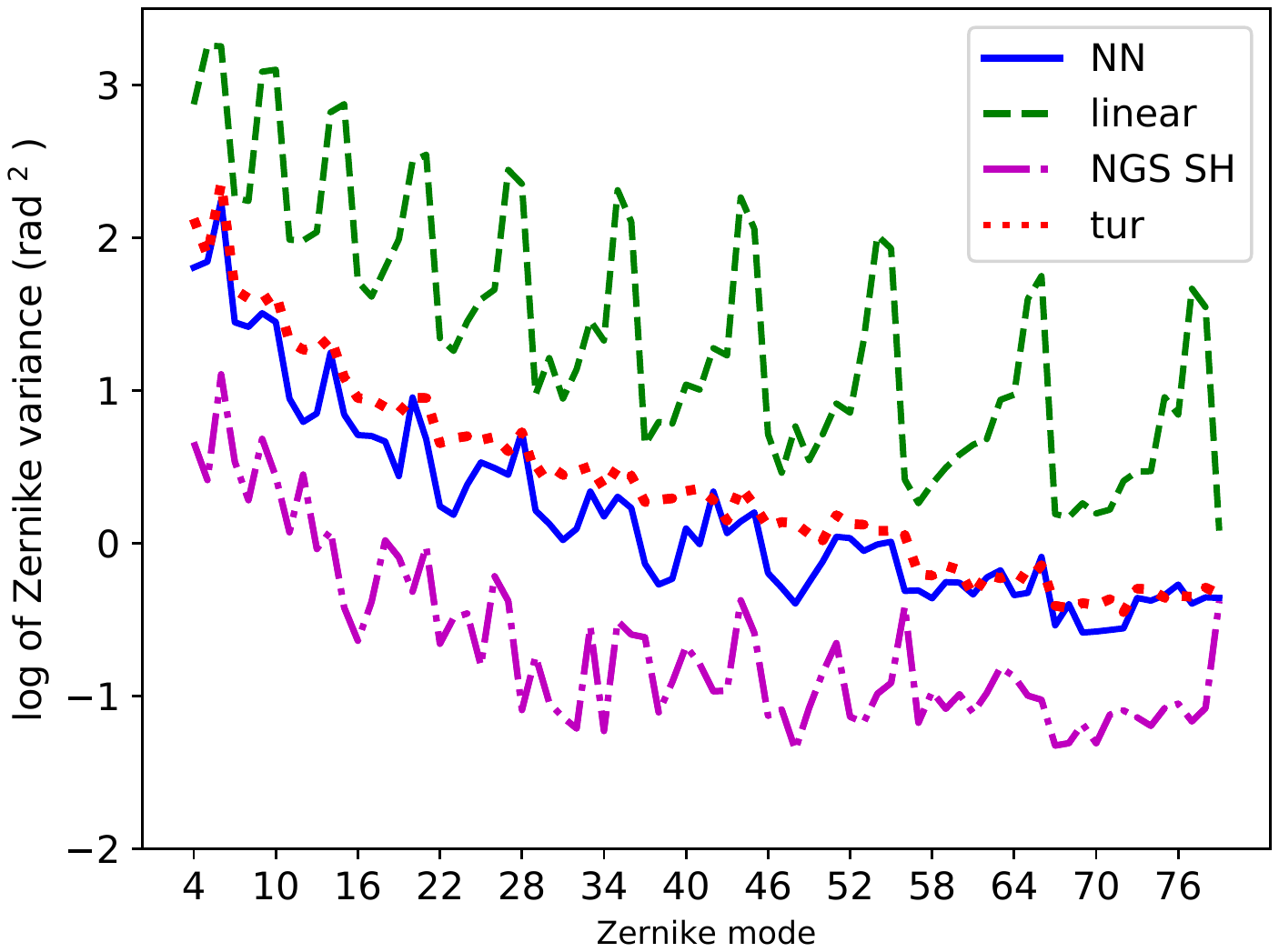}
    \label{fig:Zer_r0_025_20W}
}\\   
   \caption{Residual variance of the Zernike coefficients for the linear and NN reconstructor from AO simulation for different laser powers and for the two turbulence profiles, {\em (left)} $r_0=0.0976$ and {\em (right)} $r_0=0.171$\,m. The ``NGS SH'' lines shows the idealized performance from a noiseless SH WFS and the ``tur'' lines are the uncorrected Zernike coefficient variances. Zernike mode is synonymous with Zernike polynomial.}
    \label{fig:Zernike variance results}
    \end{centering}
\end{figure*}

To understand the source of discrepancy in reconstructor performance, the reconstructed Zernike coefficients are shown in Fig.~\ref{fig:Zernike variance results}, which shows the AO-corrected Zernike coefficients variance for laser powers equalling 20\,W, 200\,W and infinity and for both profiles. For all three wavefront sensing configurations (PPPP NN, PPPP linear and NGS SH), with no photon noise (top row) the residual is consistent with a constant fractional error. The linear PPPP retrieval, however, has a suggestion of structure consistent with smaller residuals for Zernike polynomials with smaller azimuthal frequency. This structure becomes clear when a 200\,W laser is simulated for both PPPP reconstructors, although it is weaker for the NN. For the lowest laser power, 20\,W, the correlation between the Zernike azimuthal frequency and coefficient variance becomes clear for both the PPPP reconstructors. However, the NN always gives a result with SNR$\geq1$, and therefore a useful retrieval, while the linear reconstructor has SNR$\leq1$ and is therefore a useless retrieval. In comparing the two profiles, there is some evidence that the residuals from the NN reconstructor are more closely related to the variance from the SH--which only measures $\nabla\phi$--than the linear reconstructor. Therefore it is possible that the NN preprocessing is able to enhance this signal and therefore mitigate noise effects at the potential expense of the signal otherwise available to the linear reconstructor. Such analysis to confirm this hypothesis is beyond the scope of this paper.\\

The NN reconstructor used so far is trained from the combined datasets of 10, 20, 200\,W and infinite laser power, which results in 1,200,000 independent combinations of inputs and outputs. Using this NN we demonstrate that the reconstructor has slightly worse performance (168\,nm WFE RMS for $r_0=0.0976$ and 120\,nm for $r_0=0.171$\,m) than the linear reconstructor (125\,nm WFE RMS for $r_0=0.0976$ and 86\,nm  for $r_0=0.171$\,m) for infinite laser power, but much better performance for laser powers $\leq500\,\mbox{W}$ (see Fig. \ref{fig:WFE results}).\\
The intensity of the measured beam-profile can change from laser power declining through lifetime effects or from the opacity of the atmosphere changing. The NN model used so far therefore has the advantage of being insensitive to the number of photons detected. The alternative scenario is fixing the laser power during training the NN. The result is that the performance from a single-power trained NN is only slightly better than the multiple-power trained NN but only for the specific training laser power. Table \ref{tab:WFE for different models} gives the corresponding WFEs, suggesting 160\,nm RMS when $r_0$=0.0976\,m and 125\,nm RMS when $r_0$=0.171\,m for a 4-m telescope if a 200\,W laser is used. If error sources such as the fitting and temporal errors are ignored and the tip/tilt is compensated for perfectly then the expected Strehl ratio is 0.67/0.58 in the $J$ band for $r_0=0.171$\,m when using a single/multiple power trained NN reconstructor.

\begin{table}
\begin{center}
\caption{WFE (nm) for different models using different training datasets. The first three rows use a NN trained with laser power equalling: only 20\,W, only 200\,W or a combination (10, 20 ,200\, W and infinity). The WFE of the linear reconstructor and NGS SH are shown for comparison, as well as the uncorrected turbulence RMS.}
\label{tab:WFE for different models}
\begin{tabular}{c|lcr|lcr}
\hline
\multirow{3}{*}{dataset} & \multicolumn{6}{c}{validation laser power} \\  \cline{2-7}
 & \multicolumn{3}{c|}{$r_0$=0.0976\,m} & \multicolumn{3}{c}{$r_0$=0.171\,m} \\ \cline{2-7}
 & $\infty$ & 200\,W & 20\,W  & $\infty$ & 200\,W & 20\,W \\
\hline
\hline
200\,W & 137 & 160 & 1160 & 92 & 125 & 1146\\
\hline
20\,W  & 305 & 324 & 282 & 235 & 231 & 219\\
\hline
combined  & 168 & 178 &  281 & 120 & 147 & 236 \\
\hline
\hline
linear & 125 & 248 & 1132  & 86 & 226 & 1171 \\
\hline
NGS SH & \multicolumn{3}{c|}{142} & \multicolumn{3}{c}{86} \\
\hline
\hline
Turbulence & \multicolumn{3}{c|}{460} & \multicolumn{3}{c}{290} \\
\hline
\end{tabular}
\end{center}
\end{table}

\subsection{Using one beam-profile to train the NN}
\label{sec:res:1bp}

\begin{figure}
	\begin{centering}
	\includegraphics[trim={2cm 13cm 2cm 4.7cm},clip,scale=0.5]{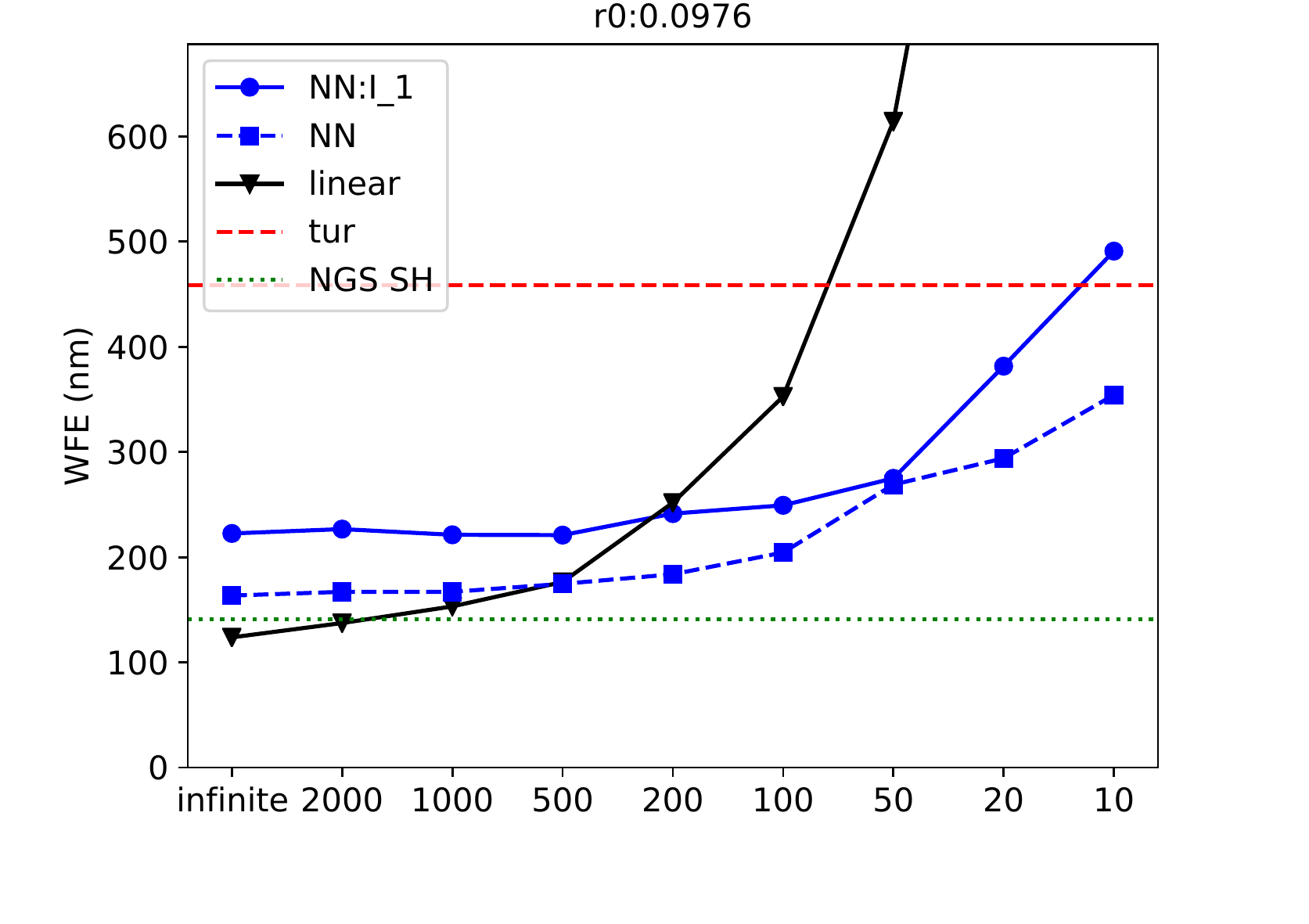}\\
    \includegraphics[trim={2cm 13cm 2cm 4.7cm},clip,scale=0.5]{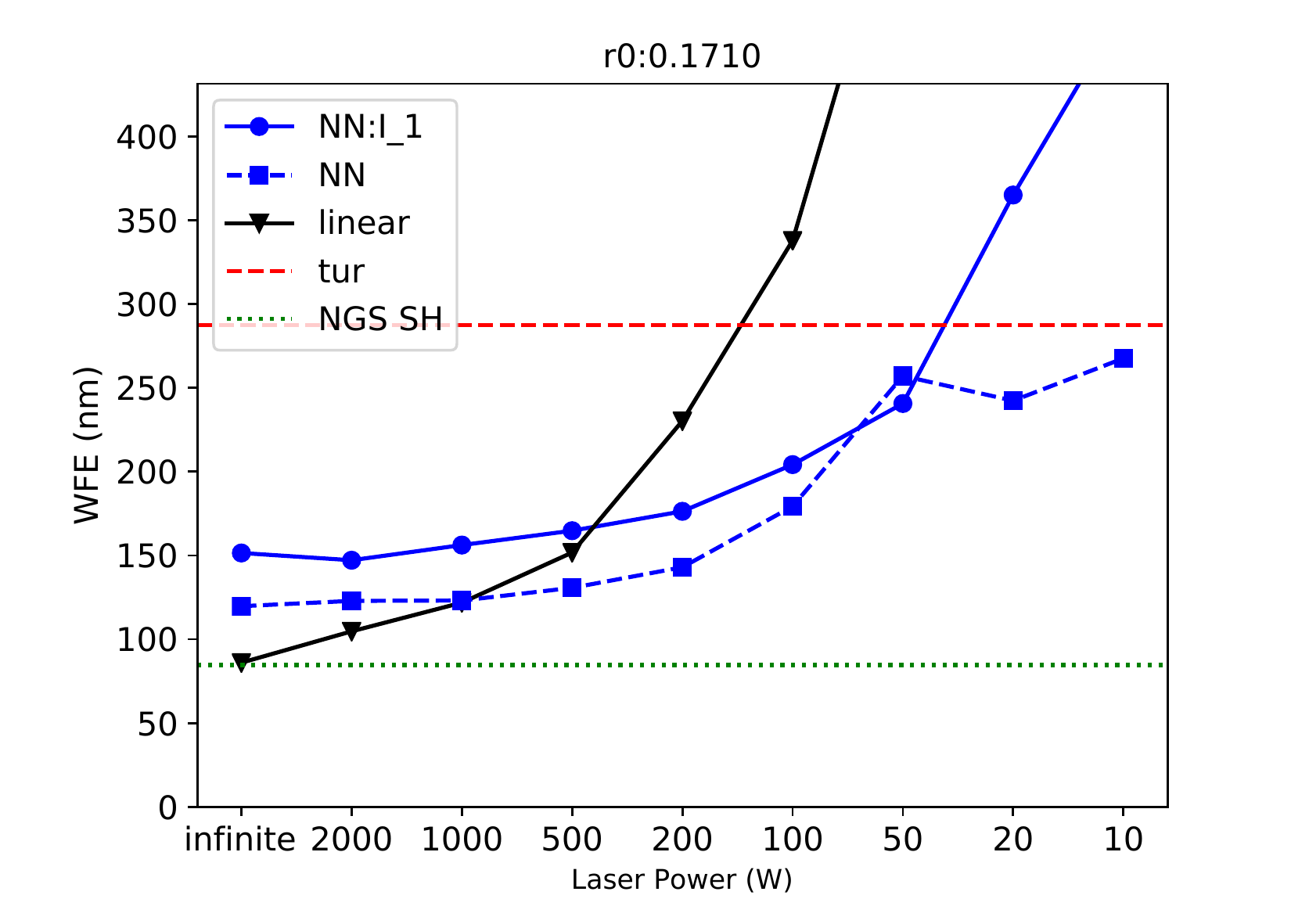}
 	\caption[]{The WFE (nm) of a $I_1$-only reconstructor. ``NN: I\_1 '' represents the NN model trained with only $I_1$ and ``NN'' represents the model trained with both $I_1$ and $I_2$ (see Fig. \ref{fig:WFE results}). The ``NGS SH'' shows the ideal performance. The result is an average of 50 random turbulence realizations from the {\em Soapy} simulations. }
    \label{fig:WFE one image}
	\end{centering}
\end{figure}

As discussed earlier, the trained NN reconstructor did not appear to use the difference of beam-profiles but instead $I_1$ and $I_2$ independently. Furthermore, a NN can be trained with just one beam-profile which suggested an experiment: can such a NN reconstructor produce a meaningful signal from turbulence distributed along the direction of laser propagation? We trained a single beam-profile NN reconstructor as described in section \ref{sec:cnn:pppp:discussion} with either $I_1$ or $I_2$ as the input component of the datasets. Both of the training datasets for 20\,W and 200\,W power were used. The corresponding results for a $I_1$-only reconstructor are shown in Fig. \ref{fig:WFE one image} (results from training with $I_2$ are worse hence not discussed further). Encouragingly, the $I_1$-only NN reconstructor shows a better performance in the simulation than the linear reconstructor--which requires both $I_1$ and $I_2$--for laser powers below 200\,W. This result points towards a simplified on-sky implementation for PPPP with a NN reconstructor wherein the camera shutter needs only be required to have an open/close repetition rate per pulse (millisecond-rates) rather than twice within a pulse (tens of microsecond-rates).

\section{Conclusions}
\label{sec:Conclusion}

We have shown that a Convolutional Neural Network can improve the performance of PPPP significantly when the laser power is below 1000\,W. Specifically we have tested the NN reconstructor using two representative turbulence profiles measured at ESO Paranal. The averaged WFEs for each profile when using the NN reconstructor are 160\,nm and 125\,nm RMS, respectively, if a 200\,W laser is used and tip/tilt is assumed perfectly compensated. These WFEs lead to J-band Strehl ratios equal to 0.52 and 0.67, respectively, compared to 0.21 and 0.28 if otherwise the linear reconstructor was used. Apart from the improved performance, another advantage of using a NN together with PPPP is that the laser beam-profile is under control unlike tomographic NGS AO, for example, which is dependent on a specific asterism for a specific target. Thus there is no need to retrain the PPPP NN reconstructor unless the laser beam-profile is changed. Considering the increased computational cost for a NN, the training process is anticipated to be offline and hence not considered here, but the real-time consideration is that the NN reconstructor requires only $\sim2\times$ more operations than the linear reconstructor. Hence the cost of using a NN reconstructor is not substantially increased. Finally, the first results of using the NN reconstructor trained with one beam-profile shows a smaller WFE for laser powers below 200\,W when compared with the linear reconstructor that demands two beam-profiles. The on-sky implementation complexity when measuring one beam-profile is significantly reduced, and only becomes possible by utilising a NN reconstructor.

\section*{Acknowledgements}

Authors from Durham University acknowledge STFC funding ST/P000541/1. Authors from Oviedo University acknowledge financial support from the I+D 2017 project AYA 2017-89121-P, and support from the European Union's Horizon 2020 research and innovation programme under the H2020-INFRAIA-2018-020 grant agreement No210489629.\\




\bibliographystyle{mnras}

\bibliography{sample} 



\bsp	
\label{lastpage}
\end{document}